\documentclass[a4paper]{article}

\usepackage{INTERSPEECH2022}
\usepackage{graphicx}
\usepackage{amsmath}
\usepackage{booktabs}
\usepackage{color,makecell,url,times, tabularx,bbm,amssymb,multirow,rotating,diagbox}
\usepackage{algorithmic,algorithm}
\usepackage{balance}

\title{RaDur: A Reference-aware and Duration-robust Network for Target Sound Detection}
\name{Dongchao Yang$^{1 \dagger}$\thanks{$^{\dagger}$ Indicates equal contribution.}, 
Helin Wang$^{1 \dagger}$, Zhongjie Ye$^1$, Yuexian Zou$^{1,*}$, Wenwu Wang$^2$\thanks{$^{*}$ Corresponding Author.}}

\address{
  $^1$ADSPLAB, School of ECE, Peking University, Shenzhen, China\\
  $^2$Center for Vision, Speech and Signal Processing, University of Surrey, UK}
\email{\{dongchao98,zhongjieye\}@stu.pku.edu.cn, \{wanghl15,zouyx\}@pku.edu.cn, w.wang@surrey.ac.uk}

\begin{document}

\maketitle
\begin{abstract}
Target sound detection (TSD) aims to detect the target sound from a mixture audio given the reference information. 
Previous methods use a conditional network to extract a sound-discriminative embedding from the reference audio, 
and then use it to detect the target sound from the mixture audio. 
However, the network performs much differently when using different reference audios 
(\text{e.g.} performs poorly for noisy and short-duration reference audios),
and tends to make wrong decisions for transient events (\text{i.e.} shorter than $1$ second). 
To overcome these problems, in this paper, we present a reference-aware and duration-robust network (RaDur) for TSD.
More specifically, in order to make the network more aware of the reference information,
we propose an embedding enhancement module   
to take into account the mixture audio while generating the embedding,
and apply the attention pooling to enhance the features of target sound-related frames and weaken the features of noisy frames.
In addition, a duration-robust focal loss is proposed to help model different-duration events. 
To evaluate our method, we build two TSD datasets based on UrbanSound and Audioset. 
Extensive experiments show the effectiveness of our methods.
\end{abstract}
\noindent\textbf{Index Terms}: target sound detection, embedding enhancement, reference-aware, duration-robust focal loss

\section{Introduction}
\label{intro}
Human beings have the ability to focus their auditory attention on a particular sound in a multi-source environment,
which attracts the related studies in machine hearing.
In this paper, we focus on the target sound detection (TSD) task \cite{yang2021detect}, 
which aims to recognize and localize target sound source within a mixture audio given a reference audio or/and a sound label,
\text{e.g.} detecting the talking sound within a noisy cafe environment.
TSD has many potential applications, 
such as noise monitoring for smart cities \cite{bello2018sonyc} and large-scale multimedia indexing \cite{hershey2017cnn}.
TSD is similar to sound event detection (SED), however, the difference is that SED aims to classify and localize all pre-defined sound events (\textit{e.g.}, train horn, car alarm) within an audio clip,
which has been widely studied \cite{dinkel2021towards,lin2020specialized,kong2020sound,mesaros2021sound, wang2019comparison,martin2019sound}. 
Other related tasks include speaker extraction \cite{wang2018voicefilter,ge2021multi,vzmolikova2019speakerbeam,borsdorf2021universal,pan2021usev} 
where the target speech is extracted from a mixture speech given a reference utterance of the target speaker,
and acoustic events sound selection (or removal) problems \cite{ochiai2020listen,delcroix2021few,okamoto2021environmental}.
Different from them, TSD focuses on the detection task, as seen in multimedia retrieval applications 
where the training data can be more easily obtained.

In a recent work \cite{yang2021detect}, a target sound detection network (TSDNet) is presented, which is composed of a conditional network and a detection network.
In TSDNet, a sound-discriminative embedding generated by the conditional network is used as the reference information
to guide the detection network for detecting the target sound from the mixture audio.
TSDNet provides a good detection performance on a small-scale training dataset (\text{i.e.} UrbanSound \cite{salamon2014dataset}).
However, 
we observe that TSDNet tends to make wrong detection on short-duration events 
(such as chop and bouncing), as Figure 2(a) shows.
In addition,
the performance of detection highly relies on the quality of the reference information,
which may be severely degraded when the reference audio is noisy or quite short.

To address these issues,
in this paper,
we propose a reference-aware and duration-robust network (called RaDur) for target sound detection task. 
More specifically, 
we design an embedding enhancement module in the conditional network, 
which utilizes the frames of mixture audio related to the reference information to enhance the embedding. 
We employ the attention pooling function to guide the conditional network to attend to target-related frames and ignore noisy frames or interference.
In addition, we apply a multi-scale feature extractor to extract characteristics of events with different duration and we propose a duration-aware focal loss to solve the problems induced by short-duration events.
To evaluate our method, we use URBAN-TSD dataset \cite{yang2021detect} and establish a new large-scale dataset (Audioset-TSD) based on Audioset \cite{hershey2017cnn}.
The experiments show that our proposed method provide 6.6\% and 16.7\% improvement for the segment-based and event-based F scores on the URBAN-TSD dataset, and 23.8\% and 8.0\% improvement for the segment-based and event-based F scores on the Audioset-TSD dataset.

\section{Proposed Method}
\label{proposed}

\begin{figure}[t]
  \centering
  \includegraphics[width=\linewidth]{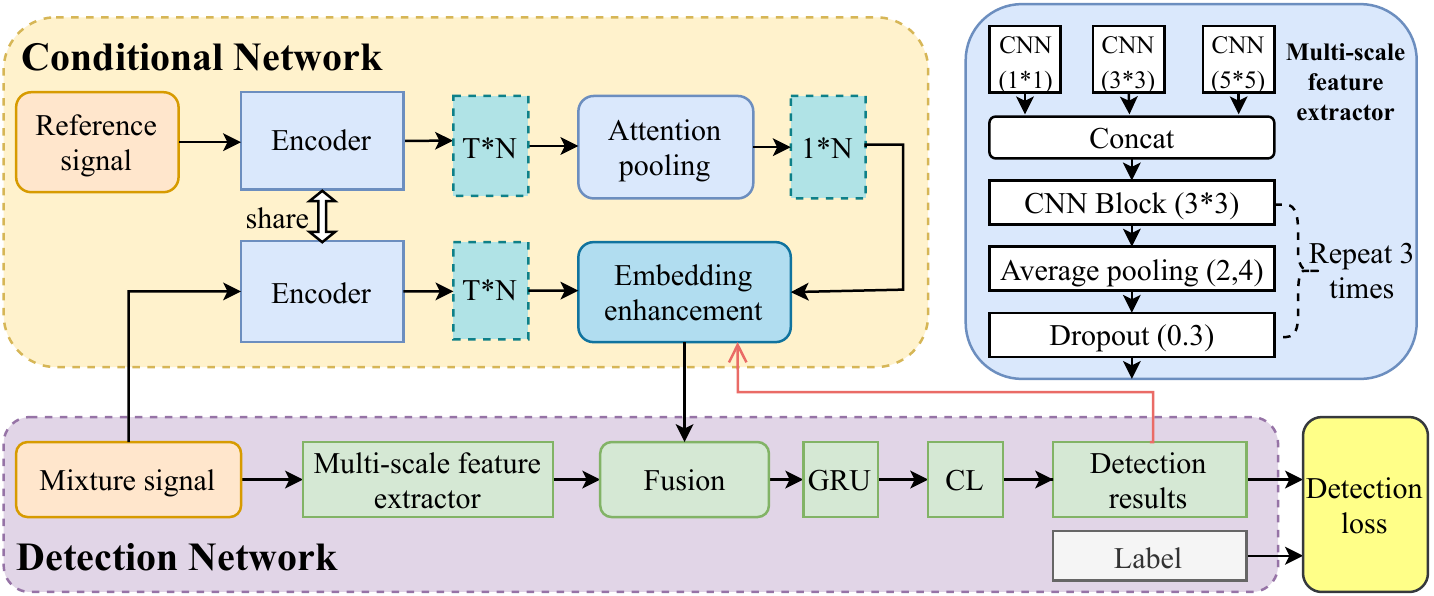}
  \caption{The architecture of our proposed RaDur. 
  Here, CL denotes the frame-level classification layer, 
  containing two fully-connected layers and one softmax function.}
  \label{fig:TEDNet+}
  \vspace*{-\baselineskip}
\end{figure}

The architecture of our proposed network (RaDur) is shown in Fig. \ref{fig:TEDNet+}, 
which is composed of two parts: a conditional network and a detection network.
In the conditional network, we propose an embedding enhancement module and use the attention pooling function to obtain more sound-discriminative embedding.
In the detection network, we apply multi-scale feature extractor to generate a multiple view of the mixture audio with respect to different-duration events.
In addition, a duration-aware focal loss function is proposed to facilitate the modelling of short-duration events. 
The details are as follows.

\subsection{Conditional Network}
The conditional network aims to extract a sound-discriminative embedding vector from the reference audio. 
Similar to the previous work \cite{yang2021detect}, 
we adopt a VGG-like convolutional neural network (CNN) model \cite{kong2020panns} for the conditional network, 
which uses the log-mel spectrogram as input and consists of $5$ convolutional blocks with $64$, $128$, $256$, $512$, $1024$ output channels, respectively. 
Lastly, we use a fully-connected layer to output the conditional embedding vector with a fixed dimension of $128$.

We observe that the quality of the embedding is crucial to TSD.
If the reference audio contains noise or if the duration of the event within the reference audio is shorter than 1 second,
the quality of the embedding and thereby the performance of TSD will be degraded. To make the embedding more discriminative and robust to noise and other interference, we propose an attention pooling function and an embedding enhancement (EE) module to enhance the embedding.

\subsubsection{Attention Pooling Function}
We find that many frames of the reference audio do not include the information of the target sound, 
instead, these frames may include noises or other interference information. 
As a result, these target-irrelevant frames may affect the distinguishability of the embedding. 
Inspired by the temporal attention pooling operations in the audio classification tasks \cite{wang2019environmental,wang2019comparison,kong2019weakly},
we replace the global pooling layer in \cite{yang2021detect}  with an attention pooling (AP) function
which enables the network to attend to the frames containing the reference information.
More specifically,
given the input representation of the reference audio $\boldsymbol{x}_{r} \in \mathcal{R}^{t_{r} \times f_{r}}$ 
where $t_r$ and $f_r$ denote the number of time frames and the number of frequency bands, respectively,
we can get the deep time-frequency feature $\boldsymbol{E}_{r} \in \mathcal{R}^{t^{\prime} \times C_{r}}$ from the encoder $f_{re}$.
\begin{equation}
\label{attention_pooling_encoder}
    \boldsymbol{E}_{r} = f_{re}(\boldsymbol{x}_{r};\theta_{r})
\end{equation}
where $t^{\prime}$ denotes the number of feature frames, 
$C_{r}$ denotes the dimension of the feature for each frame
and $\theta_{r}$ is the parameters of the encoder.
Then we use the global average pooling $f_{GAP}$ to get the global feature $\boldsymbol{e}_g \in \mathcal{R}^{C_{r}}$. 
\begin{equation}
\label{attention_pooling_GAP}
    \boldsymbol{e}_{g} = f_{GAP}(\boldsymbol{E}_{r})
\end{equation}
After that, we use the global feature $\boldsymbol{e}_g$ as query, 
$\boldsymbol{E}_{r}$ as key to calculate attention weights of all the frames $\boldsymbol{a} \in \mathcal{R}^{t^{\prime}}$.
Finally, we get the final embedding $\boldsymbol{e}_f \in \mathcal{R}^{C_{r}}$ according to the attention weights.
\begin{equation}\label{attention_pooling_attention}
    \boldsymbol{Q}=\boldsymbol{e}_g\boldsymbol{W}_q, \boldsymbol{K}=\boldsymbol{E}_r\boldsymbol{W}_k
\end{equation}
\begin{equation}\label{attention_pooling_attention}
    \boldsymbol{a} = softmax (\frac{\boldsymbol{K} \boldsymbol{Q^{T}}}{\sqrt{C_{r}}})
\end{equation}
\begin{equation}\label{attention_pooling_mul}
    \boldsymbol{E}_f = \boldsymbol{E}_r \otimes \boldsymbol{a}, \boldsymbol{e}_f = \sum_{i=1}^{t^{\prime}}E_{f}^{i}
\end{equation}
where $\boldsymbol{W}_q \in \mathcal{R}^{C_{r}\times C_{q}}$ and $\boldsymbol{W}_k \in \mathcal{R}^{C_{r}\times C_{k}}$ denote the learnable weights, and $C_{q}=C_{k}$. $\boldsymbol{E}_f = \{E_{f}^{1}, E_{f}^{2},..., E_{f}^{t^{\prime}}\} \in \mathcal{R}^{t^{\prime} \times C_{r}}$ denotes the weighted feature and $\otimes$ indicates the element-wise multiplication of a matrix and a vector.

\subsubsection{Embedding Enhancement Module}
We can mitigate the problem caused by noise and short-duration events using the attention pooling,
but the quality of some reference audios is still poor.
From the experiments, 
we find that if we use several reference audios or
directly use the target audio as the reference audio, 
we can get a much better performance. 
However, we cannot get the target audio in practical applications,
and it is sometimes hard to collect multiple audios for some unseen classes \cite{wang2020few,yang2021mutual}. 
To further improve the quality of the embedding without using extra data, 
in this paper,
we propose to use the mixture audio to enhance the embedding and present an embedding enhancement (EE) module which works at different training stages.
The core idea is to use the detection results of the previous training stages (\textit{i.e.} previous epochs in our experiments) to select frames from the features of the mixture audio that contain the characteristics of the target event. 
After that, we use these frames as the additional reference information to enhance the quality of the embedding. 

To be more specific,
we denote $f_d$ as the detection network whose inputs are the mixture audio $\boldsymbol{x}_m \in \mathcal{R}^{t_{m} \times f_{m}}$ and the embedding $\boldsymbol{e}_f$, where $t_m$ and $f_m$ denote the number of time frames and the number of frequency bands, respectively. Hence, we can get the detection results of the previous stage $\hat{\boldsymbol{y}} \in \mathcal{R}^{t^{\prime}}$ by 
\begin{equation}
\label{detection_network}
    \hat{\boldsymbol{y}} = f_{d}(\boldsymbol{x}_{m}, \boldsymbol{e}_f;\theta_{d})
\end{equation}
where $\theta_{d}$ denotes the parameters of the detection network,
and the details of the detection network are introduced in Section~\ref{detection}.
Next, we can get the deep time-frequency feature of the mixture audio $\boldsymbol{E}_{m} \in \mathcal{R}^{t^{\prime} \times C_{r}}$ from the encoder $f_{re}$ similarly to (\ref{attention_pooling_encoder}).
\begin{equation}
\label{ee_encoder}
    \boldsymbol{E}_{m} = f_{re}(\boldsymbol{x}_{m};\theta_{r})
\end{equation}
Then for the current training stage,
we select top-$k$ frames of the feature of the mixture audio according to the detection results $\hat{\boldsymbol{y}}$.
Note that $k \ll t^{\prime}$, and we can get the selected feature $\boldsymbol{E}_{m}^{\prime} \in \mathcal{R}^{k \times C_{r}}$
and the corresponding the detection scores $\hat{\boldsymbol{y}}^{\prime} \in \mathcal{R}^{k}$.
Finally,
the selected feature is used to enhance the reference embedding $\boldsymbol{e}_f$ by the attention method.
\begin{equation}\label{attention_pooling_attention}
    \boldsymbol{Q}^{\prime}=\boldsymbol{e}_f\boldsymbol{W}^{\prime}_q, \boldsymbol{K}^{\prime}=\boldsymbol{E}^{\prime}_m\boldsymbol{W}^{\prime}_k
\end{equation}
\begin{equation}\label{ee_attention}
    \boldsymbol{a}^{\prime} = softmax (\frac{\boldsymbol{K}^{\prime} {\boldsymbol{Q}^{\prime}}^{T}}{\sqrt{C_{r}}})
\end{equation}
\begin{equation}\label{ee_mul}
    \boldsymbol{E}_{s} = \boldsymbol{E}^{\prime}_m \otimes \boldsymbol{a}^{\prime}, \boldsymbol{e}_{f}^{\prime} = \sum_{i=1}^{k}E_{s}^{i}
\end{equation}
where $\boldsymbol{W}^{\prime}_q$ and $\boldsymbol{W}^{\prime}_k$ denote the learnable weights. $\boldsymbol{e}_{f}^{\prime} \in \mathcal{R}^{C_{r}}$ is the enhanced embedding,
$\boldsymbol{a}^{\prime} \in \mathcal{R}^{k}$ is the attention weights of selected frames, and
$\boldsymbol{E}_s = \{E_{s}^{1}, E_{s}^{2},..., E_{f}^{k}\} \in \mathcal{R}^{k \times C_{r}}$ denotes the weighted selected feature.
As this EE module utilizes the previous training stage to guide the current stage,
we argue that in the first several epochs, the EE module is not required, while it turns out to be important with the increase in the number of training epochs.
Thus, we set the first $10$ epochs as the warm-up stage in which the EE module is not applied, while in the following epochs, the results from the previous epoch are used to enhance the current epoch.
In addition,
if there is no target sound happening in the mixture audio,
the EE module still makes the embedding attend to the mixture audio, 
which may interfere the original reference audio.
Therefore, we set a hyper-parameter $\tau$ to control the EE module.
Here, $\tau$ is a threshold which is used to filter the detection scores $\hat{\boldsymbol{y}}^{\prime}$. Finally, we use a fusion layer (1D convolutional layer) to integrate the original embedding and the enhanced embedding.
\begin{equation}\label{ee_new_threshold}
    \hat{y}^{\prime}_{i}= \begin{cases}0, & \text { if } \hat{y}^{\prime}_{i}<\tau \\ \hat{y}^{\prime}_{i}, & \text { if } \hat{y}^{\prime}_{i} \geq \tau\end{cases}, \boldsymbol{a}^{\prime\prime} = \boldsymbol{a}^{\prime} \otimes \hat{\boldsymbol{y}}^{\prime}
\end{equation}
\begin{equation}\label{ee_new_mul}
    \boldsymbol{E}_{s}^{\prime} = \boldsymbol{E}^{\prime}_m \otimes \boldsymbol{a}^{\prime\prime},\boldsymbol{e}_{f}^{\prime\prime}=\sum_{i=1}^{k}E_{s}^{\prime^{i}}
\end{equation}
\begin{equation}\label{ee_new_add}
    \boldsymbol{e}_{f}^{\star} = \mathit{Conv1d}(\boldsymbol{e}_f) \otimes \mathit{Conv1d}(\boldsymbol{e}^{\prime\prime}_f)
\end{equation}
where $\boldsymbol{e}_{f}^{\star} \in \mathcal{R}^{C_{r}}$ is the modified enhanced embedding,
$\boldsymbol{a}^{\prime\prime} \in \mathcal{R}^{k}$ is the modified attention weights of selected frames, and
$\boldsymbol{E}_s^{\prime} = \{E_{s}^{\prime^{1}}, E_{s}^{\prime^{2}},..., E_{f}^{\prime^{k}}\} \in \mathcal{R}^{k \times C_{r}}$ denotes the modified weighted selected feature. The whole process for generating the enhanced embedding is summarized in Algorithm~\ref{alg:EE}.
\begin{algorithm}[htb]
\caption{Embedding Generation.}
\label{alg:EE}
\begin{algorithmic}[1]
\REQUIRE ~~\\
    The input representation of the reference audio $\boldsymbol{x}_r$;
    The input representation of the mixture audio $\boldsymbol{x}_m$;
\ENSURE The enhanced embedding $\boldsymbol{e}_{f}^{\star}$; \\
    \STATE Get the original embedding $\boldsymbol{e}_f$ using equation (1)-(5);
    \STATE Get the feature of the mixture audio $\boldsymbol{E}_m$ using equation (7);
    \STATE Get the detection results of the previous stage $\hat{\boldsymbol{y}}$ using equation (6);
    \STATE Select top-$k$ frames from $\boldsymbol{E}_m$, and get the selected feature $\boldsymbol{E}_{m}^{\prime}$ 
    and the corresponding detection scores $\hat{\boldsymbol{y}}$;
    \STATE Calculate the attention weights $\boldsymbol{a}^{'}$ using equation (8)-(9);
    \STATE Modify the attention weights and get $\boldsymbol{a}^{\prime\prime}$ using equation (11);
    \STATE Calculate $\boldsymbol{e}_{f}^{\star}$ according to equation (12)-(13);
\RETURN $\boldsymbol{e}_{f}^{\star}$; 
\end{algorithmic}
\end{algorithm}

\subsection{Detection Network}
\label{detection}
The detection network is composed of three parts: 
(1) Multi-scale feature extractor: 
a CNN-based structure aiming to extract the acoustic representation of the mixture audio. 
As Figure \ref{fig:TEDNet+} shows, 
to effectively capture both the global and local information which is good for the long and transient events, 
multi-scale CNNs with varied kernel sizes $1\times1$, $3\times3$ and $5\times5$ are employed \cite{xian2021multi} in the first layer
with the $64$ output channels.
Besides, we use gated linear units (GLUs) \cite{xu2018large} to replace ReLU \cite{nair2010rectified} activation in this layer. 
After that, we concatenate the output of multi-scale CNNs, 
and feed it to three CNN blocks with the output channels of $128$, $256$ and $512$ respectively to get the final representation.
(2) Bi-GRU: one Bi-GRU layer with $512$ units is used to capture temporal dependencies and integrate the conditional embedding.
(3) Frame-level classification layer: it composes of two fully-connected layers with $256$ hidden units and a softmax function, 
which is used to get the frame-level predictions.

\subsection{Loss Function}
TSDNet \cite{yang2021detect} uses the binary cross entropy (BCE) loss as the training objective.
In this paper, we find that using the focal loss \cite{lin2017focal} can get better performance especially on Audioset-TSD dataset. Furthermore, to solve the problems caused by short-duration events, a duration-aware focal loss function is proposed, which is based on the focal loss, defined by:
\begin{equation}\label{focal-loss}
    \mathcal{L}_{focal} = -\beta y(1-\hat{y})^\gamma log(\hat{y}) - (1-\beta) (1-y)\hat{y}^\gamma log(1-\hat{y})
\end{equation}
where $\beta$ and $\gamma$ are hyper-parameters.
$\hat{y}$ and $y$ denote the frame-level prediction probability and label respectively.
The focal loss is a dynamically scaled cross entropy loss, 
where the scaling factor decays to zero as confidence in the correct class increases.
Hence, the scaling factor can automatically down-weight the contributions of easy examples during
training, which allows the model to focus on hard examples.
To further improve the performance of the transient events,
we propose the duration-aware focal loss function (Du-Focal Loss),
which works by applying different weights to events of different duration.
\begin{equation}\label{duration-aware-focal-loss}
    \mathcal{L}_{Du-focal} = (1 + \alpha \frac{e^{-w}-e^{-w_{max}}}{e^{-w_{min}}-e^{-w_{max}}}) \mathcal{L}_{focal}
\end{equation}
\noindent where $w \in [0, 10]$ denotes the average duration of each event in the training dataset,
$w_{max} = 0$ is the maximum value and $w_{min} = 10$ is the minimum value.
We use $\frac{e^{-w}-e^{-w_{max}}}{e^{-w_{min}}-e^{-w_{max}}}$ to scale the values to $[0,1]$.
$\alpha$ is a hyper-parameter which controls the ratio of the extra weights of the loss.\\

\section{Audioset-TSD Dataset}
\label{dataset}
We build a new TSD dataset based on Audioset \cite{gemmeke2017audio}. 
Here, we use the strong-labelled data \cite{hershey2021benefit} 
which includes 94,126 training clips and 16,118 test clips, and they come from 456 different classes. 
We choose 192 different classes to build a large-scaled TSD dataset, named as Audioset-TSD dataset. 
The process for building the TSD dataset is the same as the previous work \cite{yang2021detect}. 
Note that Audioset-TSD dataset also includes negative samples, in which the mixture audio does not contain the target sound. 
In summary, Audioset-TSD dataset includes 10-second 490,336 training, 40,185 validation and 83,334 test clips. 
We clip the reference audio for each category directly from the Audioset training set. 
We preferentially select the audio clips that do not contain interference from other events, 
but they may still contain background noises.

As a result,
compared with URBAN-TSD dataset \cite{yang2021detect} which contains 10-class 48,489 samples, the Audioset-TSD dataset contains more data with a bigger number of classes. In addition, the clips contain noises and interference from other events. Therefore, it is a more challenging dataset.

\section{Experiments}
\label{exps}

\subsection{Experimental setups}
\noindent \textbf{Dataset.} We evaluate our methods on Audioset-TSD and URBAN-TSD \cite{yang2021detect} datasets.\\
\noindent \textbf{Metrics.} We use the segment-based F-measure and event-based F-measure \cite{mesaros2016metrics} as the evaluation metrics, which are the most commonly used metrics for sound event detection. All the F-scores are macro-averaged.\\
\textbf{Preprocessing.} The pre-extracted log mel-spectrograms with a window of 1024 samples and hop length of 320 samples are used in our experiments. The number of Mel bands is set to 64
and the size of log mel spectrogram is $1001 \times 64$. \\
\textbf{Training details.} In the training phase, the Adam \cite{kingma2014adam} is employed as the optimizer with a learning rate of $1 \times 10^{-3}$. Batch size is set to 64 and training takes 50 epochs. Following \cite{yang2021detect}, the dimension of the embedding vector is 128 and we use the multiplication fusion manner. We choose top-2 frames for the EE module.
Unless specifically stated, the hyper-parameters $\alpha$, $\beta$, $\gamma$ and $\tau$ are set to $1.5$, $0.65$, $2$ and $0.7$ respectively, empirically based on the validation set.

\begin{table}[t] \centering
\caption{The comparison between TSDNet and Radur on the URBAN-TSD dataset. S-F and E-F denote segment- and event-based F score. We did the experiments three times and report the mean value.}
\label{tab:my-table1}
\begin{tabular}{|c|c|c|c|c|c|}
\hline
Model  & Loss     & AP & EE & S-F & E-F \\ \hline
TSDNet \cite{yang2021detect} & BCE      &   &   & 69.3     & 30.6    \\ \hline
Radur  & BCE      &   &   & 71.5    &  33.3   \\ \hline
Radur  & BCE      & \checkmark  &   & 71.7     & 33.5     \\ \hline
Radur  & BCE      & \checkmark  & \checkmark  & 73.5    & 35.2    \\ \hline
Radur  & Focal    & \checkmark  & \checkmark  & \textbf{73.9}    & \textbf{35.7}    \\ \hline
Radur  & Du-Focal & \checkmark  & \checkmark  & 73.8    & \textbf{35.7}    \\ \hline
\end{tabular}
\end{table}
\begin{table}[t] \centering
\caption{The comparison between TSDNet and Radur on the Audioset-TSD dataset.}
\label{tab:my-table2}
\begin{tabular}{|c|c|c|c|c|c|}
\hline
Model  & Loss     & AP & EE & S-F & E-F \\ \hline
TSDNet \cite{yang2021detect} & BCE      &   &   & 49.5    & 48.6    \\ \hline
Radur  & BCE      &   &   & 51.3    & 50.0    \\ \hline
Radur  & BCE      & \checkmark  &   & 52.5    & 51.4    \\ \hline
Radur  & BCE      & \checkmark  & \checkmark  & 54.4    &  52.2   \\ \hline
Radur  & Focal    & \checkmark  & \checkmark  & 58.3    & 51.1    \\ \hline
Radur  & Du-Focal & \checkmark  & \checkmark  & \textbf{61.3}    &  \textbf{52.5}   \\ \hline
\end{tabular}
\end{table}

\subsection{Experimental results}
We evaluate our proposed Radur and the state-of-the-art method \cite{yang2021detect} on URBAN-TSD and Audioset-TSD datasets, and report the experimental results in Tables \ref{tab:my-table1} and \ref{tab:my-table2}. By comparing row 1 and row 2 in Tables \ref{tab:my-table1} and \ref{tab:my-table2}, we can see that RaDur has better performance thanks to the multi-scale convolutional scheme and deeper network structure. 
By comparing row 2 and row 3 in Table 2, we can see that the AP module leads to about 1.4\% improvement on the event-based F1 score, which means that the AP module can improve the quality of reference embedding. 
In addition, the comparison between row 3 and 4 shows the effectiveness of the EE module.
We can see that the focal loss performs better than the BCE loss especially on the Audioset-TSD dataset, and our proposed Du-Focal loss further improves the performance on the Audioset-TSD dataset. 
Instead, the Du-Focal loss does not bring improvement on the URBAN-TSD dataset, for the reason that all of the events in this dataset have similar duration (1-2 seconds). 

\begin{figure}[t]
  \centering
  \includegraphics[width=\linewidth]{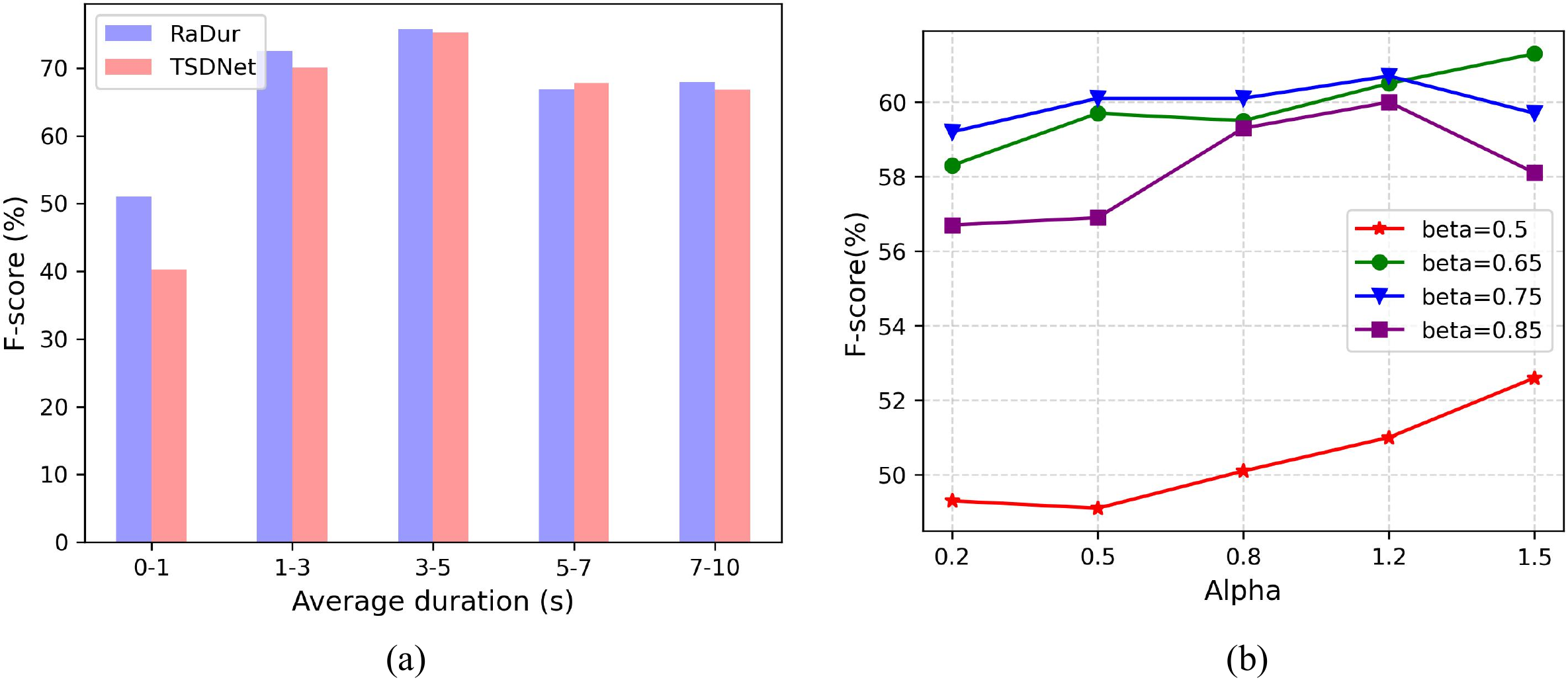}
  \caption{(a) shows the performance comparison between TSDNet and RaDur on events of different duration. (b) shows the ablation study on two hyper-parameters ($\alpha$ and $\beta$) of the Du-Focal loss. Note that F-score is segment-based.}
  \label{fig:duration}
  \vspace*{-\baselineskip}
\end{figure}

To further analyze the influence of event duration on the performance, we divide the 192 events into 5 groups according to the average duration in the test set. 
As Figure \ref{fig:duration} (a) shows, the duration of the first group is from 0 seconds to 1 seconds, which represents the transient events. The duration of the last group is from 7 seconds to 10 seconds, which represents the long events. We can find that detecting transient events is the most difficult task. Because of the multi-scale feature extractor and duration-aware focal loss, RaDur provides significant improvement on transient events. However, the detection results for transient and long events still have a large gap, which deserves further study in our future work.\\
\subsection{Ablation Studies}
We first conduct ablation studies to investigate the effect of hyper-parameter $\tau$ on the EE module and report experimental results in Table 3. We can see that the performance gradually improves with $\tau$ increasing from 0.5 to 0.7. After that, the performance begins to decrease. The experimental phenomenon meets our hypothesis: If $\tau$ is smaller than 0.6, we may select non-target feature frame, which may influence the original embedding. On the contrary, if $\tau$ is larger that 0.8, most of the frames will be filtered, then the EE module may not work. 

We also conduct ablation studies to investigate the influence of two of the hyper-parameters ($\alpha$ and $\beta$) in the Du-Focal loss.
As shown in Figure 2 (b), the value of $\beta$ is very important, and $\beta$ should be larger than 0.5 because there are more negative sample frames than positive ones in the Audioset-TSD dataset. 
Furthermore, as the value of $\alpha$ increases, the F-score is boosted. But we can see that if $\beta > 0.65$ and $\alpha>1.2$, the performance will drop. We argue that if $\beta$ and $\alpha$ are both set too large, the ratios of negative samples will be too small, as a result, the negative samples may be easily ignored in the training process.

\begin{table}[t] \centering
\caption{Ablation studies on hyper-parameter $\tau$ on the URBAN-TSD dataset.}
\label{tab:my-table3}
\begin{tabular}{|c|c|c|c|}
\hline
Model & $\tau$ & \begin{tabular}[c]{@{}c@{}}Segment-based\\ F-score\end{tabular} & \begin{tabular}[c]{@{}c@{}}Event-based\\ F-score\end{tabular} \\ \hline
Radur & 0.5                 & 71.4                                                                &  32.9                                                             \\ \hline
Radur & 0.6                 & 72.6                                                                &  34.1                                                             \\ \hline
Radur & 0.7                 & \textbf{73.5}                                                                &  \textbf{35.2}                                                             \\ \hline
Radur & 0.8                 & 73.1                                                                &  34.9                                                             \\ \hline
Radur & 0.9                 & 71.6                                                                &  33.8                                                             \\ \hline
\end{tabular}
\end{table}

\section{Conclusions}
\label{cons}
We have presented an improved TSDNet (RaDur) by modelling short-duration events and enhancing the discriminating ability of the embedding vectors. In the future work,
we will explore the extension of novel classes. The source code and dateset of this work have been released.\footnote{https://github.com/yangdongchao/RaDur}
\section{Acknowledgements}
This paper was supported by Shenzhen Science and Technology Fundamental Research Programs JSGG20191129105421211 and GXWD20201231165807007-20200814115301001.

\bibliographystyle{IEEEtran}

\bibliography{mybib}

\begin{thebibliography}{10}
\providecommand{\url}[1]{#1}
\csname url@samestyle\endcsname
\providecommand{\newblock}{\relax}
\providecommand{\bibinfo}[2]{#2}
\providecommand{\BIBentrySTDinterwordspacing}{\spaceskip=0pt\relax}
\providecommand{\BIBentryALTinterwordstretchfactor}{4}
\providecommand{\BIBentryALTinterwordspacing}{\spaceskip=\fontdimen2\font plus
\BIBentryALTinterwordstretchfactor\fontdimen3\font minus
  \fontdimen4\font\relax}
\providecommand{\BIBforeignlanguage}[2]{{%
\expandafter\ifx\csname l@#1\endcsname\relax
\typeout{** WARNING: IEEEtran.bst: No hyphenation pattern has been}%
\typeout{** loaded for the language `#1'. Using the pattern for}%
\typeout{** the default language instead.}%
\else
\language=\csname l@#1\endcsname
\fi
#2}}
\providecommand{\BIBdecl}{\relax}
\BIBdecl

\bibitem{yang2021detect}
D.~Yang, H.~Wang, Y.~Zou, and C.~Weng, ``Detect what you want: Target sound
  detection,'' \emph{arXiv preprint arXiv:2112.10153}, 2021.

\bibitem{bello2018sonyc}
J.~P. Bello, C.~Silva, O.~Nov, R.~DuBois, A.~Arora, J.~Salamon, C.~Mydlarz, and
  H.~Doraiswamy, ``Sonyc: A system for the monitoring, analysis and mitigation
  of urban noise pollution,'' \emph{arXiv preprint arXiv:1805.00889}, 2018.

\bibitem{hershey2017cnn}
S.~Hershey, S.~Chaudhuri, D.~Ellis, J.~Gemmeke, A.~Jansen, R.~C. Moore,
  M.~Plakal, D.~Platt, R.~A. Saurous, B.~Seybold \emph{et~al.}, ``{CNN}
  architectures for large-scale audio classification,'' in \emph{IEEE
  International Conference on Acoustics, Speech and Signal Processing
  (ICASSP)}.\hskip 1em plus 0.5em minus 0.4em\relax IEEE, 2017, pp. 131--135.

\bibitem{dinkel2021towards}
H.~Dinkel, M.~Wu, and K.~Yu, ``Towards duration robust weakly supervised sound
  event detection,'' \emph{IEEE/ACM Transactions on Audio, Speech, and Language
  Processing}, vol.~29, pp. 887--900, 2021.

\bibitem{lin2020specialized}
L.~Lin, X.~Wang, H.~Liu, and Y.~Qian, ``Specialized decision surface and
  disentangled feature for weakly-supervised polyphonic sound event
  detection,'' \emph{IEEE/ACM Transactions on Audio, Speech, and Language
  Processing}, vol.~28, pp. 1466--1478, 2020.

\bibitem{kong2020sound}
Q.~Kong, Y.~Xu, W.~Wang, and M.~D. Plumbley, ``Sound event detection of weakly
  labelled data with cnn-transformer and automatic threshold optimization,''
  \emph{IEEE/ACM Transactions on Audio, Speech, and Language Processing},
  vol.~28, pp. 2450--2460, 2020.

\bibitem{mesaros2021sound}
A.~Mesaros, T.~Heittola, T.~Virtanen, and M.~D. Plumbley, ``Sound event
  detection: A tutorial,'' \emph{IEEE Signal Processing Magazine}, vol.~38,
  no.~5, pp. 67--83, 2021.

\bibitem{wang2019comparison}
Y.~Wang, J.~Li, and F.~Metze, ``A comparison of five multiple instance learning
  pooling functions for sound event detection with weak labeling,'' in
  \emph{IEEE International Conference on Acoustics, Speech and Signal
  Processing (ICASSP)}.\hskip 1em plus 0.5em minus 0.4em\relax IEEE, 2019, pp.
  31--35.

\bibitem{martin2019sound}
I.~Mart{\'\i}n-Morat{\'o}, A.~Mesaros, T.~Heittola, T.~Virtanen, M.~Cobos, and
  F.~Ferri, ``Sound event envelope estimation in polyphonic mixtures,'' in
  \emph{IEEE International Conference on Acoustics, Speech and Signal
  Processing (ICASSP)}.\hskip 1em plus 0.5em minus 0.4em\relax IEEE, 2019, pp.
  935--939.

\bibitem{wang2018voicefilter}
Q.~Wang, H.~Muckenhirn, K.~Wilson, P.~Sridhar, Z.~Wu, J.~R. Hershey, R.~A.
  Saurous, R.~J. Weiss, Y.~Jia, and I.~L. Moreno, ``Voicefilter: Targeted voice
  separation by speaker-conditioned spectrogram masking,'' \emph{Proc.
  Interspeech}, pp. 2728--2732, 2019.

\bibitem{ge2021multi}
M.~Ge, C.~Xu, L.~Wang, E.~S. Chng, J.~Dang, and H.~Li, ``Multi-stage speaker
  extraction with utterance and frame-level reference signals,'' in \emph{IEEE
  International Conference on Acoustics, Speech and Signal Processing
  (ICASSP)}.\hskip 1em plus 0.5em minus 0.4em\relax IEEE, 2021, pp. 6109--6113.

\bibitem{vzmolikova2019speakerbeam}
K.~{\v{Z}}mol{\'\i}kov{\'a}, M.~Delcroix, K.~Kinoshita, T.~Ochiai, T.~Nakatani,
  L.~Burget, and J.~{\v{C}}ernock{\`y}, ``Speakerbeam: Speaker aware neural
  network for target speaker extraction in speech mixtures,'' \emph{IEEE
  Journal of Selected Topics in Signal Processing}, vol.~13, no.~4, pp.
  800--814, 2019.

\bibitem{borsdorf2021universal}
M.~Borsdorf, C.~Xu, H.~Li, and T.~Schultz, ``Universal speaker extraction in
  the presence and absence of target speakers for speech of one and two
  talkers,'' in \emph{Proc. Interspeech}, 2021, pp. 1469--1473.

\bibitem{pan2021usev}
Z.~Pan, M.~Ge, and H.~Li, ``Usev: Universal speaker extraction with visual
  cue,'' \emph{arXiv preprint arXiv:2109.14831}, 2021.

\bibitem{ochiai2020listen}
T.~Ochiai, M.~Delcroix, Y.~Koizumi, H.~Ito, K.~Kinoshita, and S.~Araki,
  ``Listen to what you want: Neural network-based universal sound selector,''
  \emph{Proc. Interspeech}, pp. 1441--1445, 2020.

\bibitem{delcroix2021few}
M.~Delcroix, J.~B. V{\'a}zquez, T.~Ochiai, K.~Kinoshita, and S.~Araki,
  ``Few-shot learning of new sound classes for target sound extraction,''
  \emph{arXiv preprint arXiv:2106.07144}, 2021.

\bibitem{okamoto2021environmental}
Y.~Okamoto, S.~Horiguchi, M.~Yamamoto, K.~Imoto, and Y.~Kawaguchi,
  ``Environmental sound extraction using onomatopoeia,'' \emph{arXiv preprint
  arXiv:2112.00209}, 2021.

\bibitem{salamon2014dataset}
J.~Salamon, C.~Jacoby, and J.~P. Bello, ``A dataset and taxonomy for urban
  sound research,'' in \emph{Proceedings of the 22nd ACM International
  Conference on Multimedia}, 2014, pp. 1041--1044.

\bibitem{kong2020panns}
Q.~Kong, Y.~Cao, T.~Iqbal, Y.~Wang, W.~Wang, and M.~Plumbley, ``Panns:
  Large-scale pretrained audio neural networks for audio pattern recognition,''
  \emph{IEEE/ACM Transactions on Audio, Speech, and Language Processing},
  vol.~28, pp. 2880--2894, 2020.

\bibitem{wang2019environmental}
H.~Wang, Y.~Zou, D.~Chong, and W.~Wang, ``Environmental sound classification
  with parallel temporal-spectral attention,'' in \emph{Proc. Interspeech},
  2020, pp. 821--825.

\bibitem{kong2019weakly}
Q.~Kong, C.~Yu, Y.~Xu, T.~Iqbal, W.~Wang, and M.~D. Plumbley, ``Weakly labelled
  audioset tagging with attention neural networks,'' \emph{IEEE/ACM
  Transactions on Audio, Speech, and Language Processing}, vol.~27, no.~11, pp.
  1791--1802, 2019.

\bibitem{wang2020few}
Y.~Wang, J.~Salamon, N.~J. Bryan, and J.~P. Bello, ``Few-shot sound event
  detection,'' in \emph{IEEE International Conference on Acoustics, Speech and
  Signal Processing (ICASSP)}.\hskip 1em plus 0.5em minus 0.4em\relax IEEE,
  2020, pp. 81--85.

\bibitem{yang2021mutual}
D.~Yang, H.~Wang, Y.~Zou, Z.~Ye, and W.~Wang, ``A mutual learning framework for
  few-shot sound event detection,'' \emph{arXiv preprint arXiv:2110.04474},
  2021.

\bibitem{xian2021multi}
Y.~Xian, Y.~Sun, W.~Wang, and S.~M. Naqvi, ``Multi-scale residual convolutional
  encoder decoder with bidirectional long short-term memory for single channel
  speech enhancement,'' in \emph{IEEE European Signal Processing Conference
  (EUSIPCO)}.\hskip 1em plus 0.5em minus 0.4em\relax IEEE, 2021, pp. 431--435.

\bibitem{xu2018large}
Y.~Xu, Q.~Kong, W.~Wang, and M.~D. Plumbley, ``Large-scale weakly supervised
  audio classification using gated convolutional neural network,'' in
  \emph{International Conference on Acoustics, Speech and Signal Processing
  (ICASSP)}.\hskip 1em plus 0.5em minus 0.4em\relax IEEE, 2018, pp. 121--125.

\bibitem{nair2010rectified}
V.~Nair and G.~Hinton, ``Rectified linear units improve restricted boltzmann
  machines,'' in \emph{ICML}, 2010.

\bibitem{lin2017focal}
T.-Y. Lin, P.~Goyal, R.~Girshick, K.~He, and P.~Doll{\'a}r, ``Focal loss for
  dense object detection,'' in \emph{Proceedings of the IEEE International
  Conference on Computer Vision}, 2017, pp. 2980--2988.

\bibitem{gemmeke2017audio}
J.~Gemmeke, D.~Ellis, D.~Freedman, A.~Jansen, W.~Lawrence, R.~C. Moore,
  M.~Plakal, and M.~Ritter, ``Audio set: An ontology and human-labeled dataset
  for audio events,'' in \emph{IEEE International Conference on Acoustics,
  Speech and Signal Processing (ICASSP)}.\hskip 1em plus 0.5em minus
  0.4em\relax IEEE, 2017, pp. 776--780.

\bibitem{hershey2021benefit}
S.~Hershey, D.~P. Ellis, E.~Fonseca, A.~Jansen, C.~Liu, R.~C. Moore, and
  M.~Plakal, ``The benefit of temporally-strong labels in audio event
  classification,'' in \emph{IEEE International Conference on Acoustics, Speech
  and Signal Processing (ICASSP)}.\hskip 1em plus 0.5em minus 0.4em\relax IEEE,
  2021, pp. 366--370.

\bibitem{mesaros2016metrics}
A.~Mesaros, T.~Heittola, and T.~Virtanen, ``Metrics for polyphonic sound event
  detection,'' \emph{Applied Sciences}, vol.~6, no.~6, p. 162, 2016.

\bibitem{kingma2014adam}
D.~P. Kingma and J.~Ba, ``Adam: A method for stochastic optimization,'' in
  \emph{International Conference for Learning Representations (ICML)}, 2015.

\end{thebibliography}
\end{document}